\documentclass[journal=jctcce,manuscript=article]{achemso}
\setkeys{acs}{articletitle = true, chaptertitle = true}

\usepackage[version=3]{mhchem}
\usepackage{multirow}
\usepackage{amsmath}
\usepackage{atbegshi}
\usepackage{listings}
\usepackage{bm}
\usepackage{algorithm}
\AtBeginDocument{\addtocounter{page}{-1}\AtBeginShipoutNext{\AtBeginShipoutDiscard}}

\author{Jinzhe Zeng}
\affiliation{School of Artificial Intelligence and Data Science, Unversity of Science and Technology of China, Hefei, P. R.~China}
\email{jinzhe.zeng@ustc.edu.cn} 
\author{Duo Zhang}
\affiliation{AI for Science Institute, Beijing 100080, P. R.~China}
\alsoaffiliation{DP Technology, Beijing 100080, P. R.~China}
\alsoaffiliation{Academy for Advanced Interdisciplinary Studies, Peking University, Beijing 100871, P. R.~China}
\author{Anyang Peng}
\affiliation{AI for Science Institute, Beijing 100080, P. R.~China}
\author{Xiangyu Zhang}
\affiliation{State Key Lab of Processors, Institute of Computing Technology, Chinese Academy of Sciences, Beijing 100871, P.R.~China}
\alsoaffiliation{University of Chinese Academy of Sciences, Beijing, P. R.~China}
\author{Sensen He}
\affiliation{Baidu Inc., Beijing, P. R.~China}
\author{Yan Wang}
\affiliation{State Key Lab of Processors, Institute of Computing Technology, Chinese Academy of Sciences, Beijing 100871, P.R.~China}
\alsoaffiliation{University of Chinese Academy of Sciences, Beijing, P. R.~China}
\author{Xinzijian Liu}
\affiliation{DP Technology, Beijing 100080, P. R.~China}
\author{Hangrui Bi}
\affiliation{Department of Computer Science, University of Toronto, Toronto, Ontario, Canada}
\author{Yifan Li}
\affiliation{Department of Chemistry, Princeton University, Princeton, New Jersey 08540, United States}
\author{Chun Cai}
\affiliation{AI for Science Institute, Beijing 100080, P. R.~China}
\author{Chengqian Zhang}
\affiliation{Academy for Advanced Interdisciplinary Studies, Peking University, Beijing 100871, P. R.~China}
\author{Yiming Du}
\affiliation{State Key Lab of Processors, Institute of Computing Technology, Chinese Academy of Sciences, Beijing 100871, P.R.~China}
\alsoaffiliation{University of Chinese Academy of Sciences, Beijing 100871, P.R.~China}
\author{Jia-Xin Zhu}
\affiliation{State Key Laboratory of Physical Chemistry of Solid Surfaces, iChEM, College of Chemistry and Chemical Engineering, Xiamen University, Xiamen 361005, P.R.~China}
\author{Pinghui Mo}
\affiliation{College of Integrated Circuits, Hunan University, Changsha, 410082, P.R.~China}
\author{Zhengtao Huang}
\affiliation{State Key Laboratory of Advanced Technology for Materials Synthesis and Processing, Center for Smart Materials and Device Integration, School of Material Science and Engineering, Wuhan University of Technology, Wuhan, 430070, P.R.~China}
\author{Qiyu Zeng}
\affiliation{College of Science, National University of Defense Technology, Changsha 410073, P.R.~China}
\alsoaffiliation{Hunan Key Laboratory of Extreme Matter and Applications, National University of Defense Technology, Changsha 410073, P.R.~China}
\author{Shaochen Shi}
\affiliation{ByteDance Research, Beijing 100098, P. R. China}
\author{Xuejian Qin}
\affiliation{Ningbo Institute of Materials Technology and Engineering, Chinese Academy of Sciences, Ningbo 315201, P.R.~China}
\alsoaffiliation{College of Materials Science and Opto-Electronic Technology, University of Chinese Academy of Sciences, Beijing 100049, P.R.~China}
\author{Zhaoxi Yu}
\affiliation{Key Laboratory of Theoretical and Computational Photochemistry of Ministry of Education, College of Chemistry, Beijing Normal University, Beijing 100875, P. R.~China}
\author{Chenxing Luo}
\affiliation{Department of Geosciences, Princeton University, Princeton, New Jersey 08544, United States}
\alsoaffiliation{Department of Applied Physics and Applied Mathematics, Columbia University, New York, New York 10027, United States}
\author{Ye Ding}
\affiliation{DP Technology, Beijing 100080, P. R.~China}
\author{Yun-Pei Liu}
\affiliation{Laboratory of AI for Electrochemistry (AI4EC), IKKEM, Xiamen, 361005, Fujian, P. R.~China}
\author{Ruosong Shi}
\affiliation{Graduate School of China Academy of Engineering Physics, Beijing 100088, P. R.~China}
\author{Zhenyu Wang}
\affiliation{Key Laboratory of Material Simulation Methods and Software of Ministry of Education, College of Physics, Jilin University, Changchun 130012, P.R.~China}
\alsoaffiliation{International Center of Future Science, Jilin University, Changchun, 130012, P.R.~China}
\author{Sigbjørn Løland Bore}
\affiliation{Department of Chemistry and Hylleraas Centre for Quantum Molecular Sciences, University of Oslo, 0315 Oslo, Norway}
\author{Junhan Chang}
\affiliation{DP Technology, Beijing 100080, P. R.~China}
\alsoaffiliation{College of Chemistry and Molecular Engineering, Peking University, Beijing 100871, P. R.~China}
\author{Zhe Deng}
\affiliation{College of Chemistry and Molecular Engineering, Peking University, Beijing 100871, P. R.~China}
\author{Zhaohan Ding}
\affiliation{DP Technology, Beijing 100080, P. R.~China}
\author{Siyuan Han}
\affiliation{New Cornerstone Science Laboratory, State Key Laboratory for Physical Chemistry of Solid Surfaces, Collaborative Innovation Center of Chemistry for Energy Materials, and College of Chemistry and Chemical Engineering, Xiamen University, Xiamen 361005, P. R.~China}
\author{Wanrun Jiang}
\affiliation{AI for Science Institute, Beijing 100080, P. R.~China}
\author{Guolin Ke}
\affiliation{DP Technology, Beijing 100080, P. R.~China}
\author{Zhaoqing Liu}
\affiliation{College of Chemistry and Molecular Engineering, Peking University, Beijing 100871, P. R.~China}
\author{Denghui Lu}
\affiliation{Department of Mechanics and Engineering Science, and HEDPS and CAPT, College of Engineering, Peking University, Beijing 100871, P. R.~China}
\author{Koki Muraoka}
\affiliation{Department of Chemical System Engineering, The University of Tokyo, 7-3-1 Hongo, Bunkyo-ku, Tokyo 113-8656, Japan}
\author{Hananeh Oliaei}
\affiliation{Department of Mechanical Science and Engineering, University of Illinois at Urbana-Champaign, Urbana, IL 61801, United States}
\author{Anurag Kumar Singh}
\affiliation{Department of Data Science, Indian Institute of Technology Palakkad, Kerala, India}
\author{Haohui Que}
\affiliation{Shanghai Astronomical Observatory,Chinese Academy of Sciences, Shanghai, P.R.~China}
\author{Weihong Xu}
\affiliation{Laboratory of AI for Electrochemistry (AI4EC), IKKEM, Xiamen, 361005, Fujian, P. R.~China}
\author{Zhangmancang Xu}
\affiliation{International School of Materials Science and Engineering, Wuhan University of Technology, Wuhan 430070, P. R.~China}
\author{Yong-Bin Zhuang}
\affiliation{Chaire de Simulation \`a l'Echelle Atomique (CSEA), Ecole Polytechnique F\'ed\'erale de Lausanne (EPFL), Lausanne CH-1015, Switzerland}
\author{Jiayu Dai}
\affiliation{College of Science, National University of Defense Technology, Changsha 410073, P.R.~China}
\alsoaffiliation{Hunan Key Laboratory of Extreme Matter and Applications, National University of Defense Technology, Changsha 410073, P.R.~China}
\author{Timothy J. Giese}
\affiliation{Laboratory for Biomolecular Simulation Research, Institute for Quantitative Biomedicine and Department of Chemistry and Chemical Biology, Rutgers University, Piscataway, NJ 08854, USA}
\author{Weile Jia}
\affiliation{State Key Lab of Processors, Institute of Computing Technology, Chinese Academy of Sciences, Beijing 100871, P.R.~China}
\author{Ben Xu}
\affiliation{Graduate School of Chinese Academy of Engineering Physics, Beijing, China}
\author{Darrin M. York}
\affiliation{Laboratory for Biomolecular Simulation Research, Institute for Quantitative Biomedicine and Department of Chemistry and Chemical Biology, Rutgers University, Piscataway, NJ 08854, USA}
\author{Linfeng Zhang}
\affiliation{AI for Science Institute, Beijing 100080, P. R.~China}
\alsoaffiliation{DP Technology, Beijing 100080, P. R.~China}
\email{linfeng.zhang.zlf@gmail.com}
\author{Han Wang}
\affiliation{National Key Laboratory of Computational Physics, Institute of Applied Physics and Computational Mathematics, Fenghao East Road 2, Beijing 100094, P.R.~China}
\alsoaffiliation{HEDPS, CAPT, College of Engineering, Peking University, Beijing 100871, P.R.~China}
\email{wang_han@iapcm.ac.cn}

\title{DeePMD-kit v3: A Multiple-Backend Framework for Machine Learning Potentials}

\begin{document}

\begin{tocentry}
  \includegraphics[width=\linewidth]{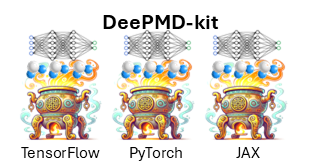}
\end{tocentry}

\begin{abstract}
In recent years, machine learning potentials (MLPs) have become indispensable tools in physics, chemistry, and materials science, driving the development of software packages for molecular dynamics (MD) simulations and related applications.
These packages, typically built on specific machine learning frameworks such as TensorFlow, PyTorch, or JAX, face integration challenges when advanced applications demand communication across different frameworks.
The previous TensorFlow-based implementation of DeePMD-kit exemplified these limitations.
In this work, we introduce DeePMD-kit version 3, a significant update featuring a multi-backend framework that supports TensorFlow, PyTorch, JAX, and PaddlePaddle backends, and demonstrate the versatility of this architecture through the integration of other MLPs packages and of Differentiable Molecular Force Field.
This architecture allows seamless backend switching with minimal modifications, enabling users and developers to integrate DeePMD-kit with other packages using different machine learning frameworks. This innovation facilitates the development of more complex and interoperable workflows, paving the way for broader applications of MLPs in scientific research.
\end{abstract}

\section{Introduction}
Over the past decade, machine learning potentials (MLPs) have become increasingly influential in the fields of physics, chemistry, molecular biology, and materials science.
\cite{Behler_JChemPhys_2016_v145_p170901,Butler_Nature_2018_v559_p547,Noe_AnnuRevPhysChem_2020_v71_p361,Pinheiro_ChemSci_2021_v12_p14396,Manzhos_ChemRev_2021_v121_p10187,Zeng_BookChap_QuantChemML_2022_p279}
This has led to the development of software packages specifically designed for training and employing MLPs in molecular dynamics (MD) and free energy simulations, and other applications requiring accurate potential energy and force calculations.
\cite{Wang_ComputPhysCommun_2018_v228_p178,Schutt_JChemTheoryComput_2019_v15_p448,Chmiela_ComputPhysCommun_2019_v240_p38,Unke_JChemTheoryComput_2019_v15_p3678,Lee_ComputPhysCommun_2019_v242_p95,Gao_JChemInfModel_2020_v60_p3408,Dral_TopCurrChem_2021_v379_p27,Singraber_JChemTheoryComput_2019_v15_p1827,Zhang_JChemPhys_2022_v156_p114801,Schutt_JChemPhys_2023_v158_p144801,Fan_JChemPhys_2022_v157_p114801,Novikov_MachLearnSciTechnol_2021_v2_p025002,Yanxon_MachLearnSciTechnol_2021_v2_p027001,Zeng_JChemTheoryComput_2023_v19_p1261,Zeng_JChemPhys_2023_v158_p124110}
These packages are typically built upon specific machine learning frameworks such as TensorFlow\cite{Abadi_2015_tensorflow}, PyTorch\cite{Paszke_arXiv_2019_p1912.01703}, or JAX\cite{jax2018github}.
While a single machine learning framework often meets most requirements, increasingly complex applications necessitate interoperability between packages utilizing different machine learning frameworks.
This poses a significant challenge for developers when these packages rely on disparate frameworks.
Additionally, each framework may have unique advantages, such as optimized performance for specific applications or better compatibility with certain hardware.
To address these limitations, developers of several MLP packages have created secondary versions based on alternative frameworks.
For instance, MACE\cite{Batatia_BookChap_NIPS_2022_v36_p830} offers both PyTorch-based and JAX-based implementations.

The earlier version of the DeePMD-kit package\cite{Wang_ComputPhysCommun_2018_v228_p178,Zeng_JChemPhys_2023_v159_p054801} was built on TensorFlow, which posed challenges for integration with other packages utilizing different deep-learning frameworks.
To address this limitation, Gao \textit{et al.} developed a JAX-based package \cite{Gao_PhysChemChemPhys_2024_v26_p23080} to enable seamless integration of deep potential models with JAX-MD\cite{Schoenholz_JStatMech_2021_v2021_p124016} for an end-to-end GPU-accelerated workflow.
Similarly, Zhang \textit{et al.} introduced a PyTorch-based package alongside the development of the DPA-2 model\cite{Zhang_npjComputMater_2024_v10_p293} for better distributed training performance.
However, maintaining consistent user interfaces (UIs) and application programming interfaces (APIs) for tasks such as training, inference, and molecular dynamics simulations across these different frameworks proved to be both inconvenient and inefficient.

\begin{figure}[t]
    \includegraphics[width=\linewidth]{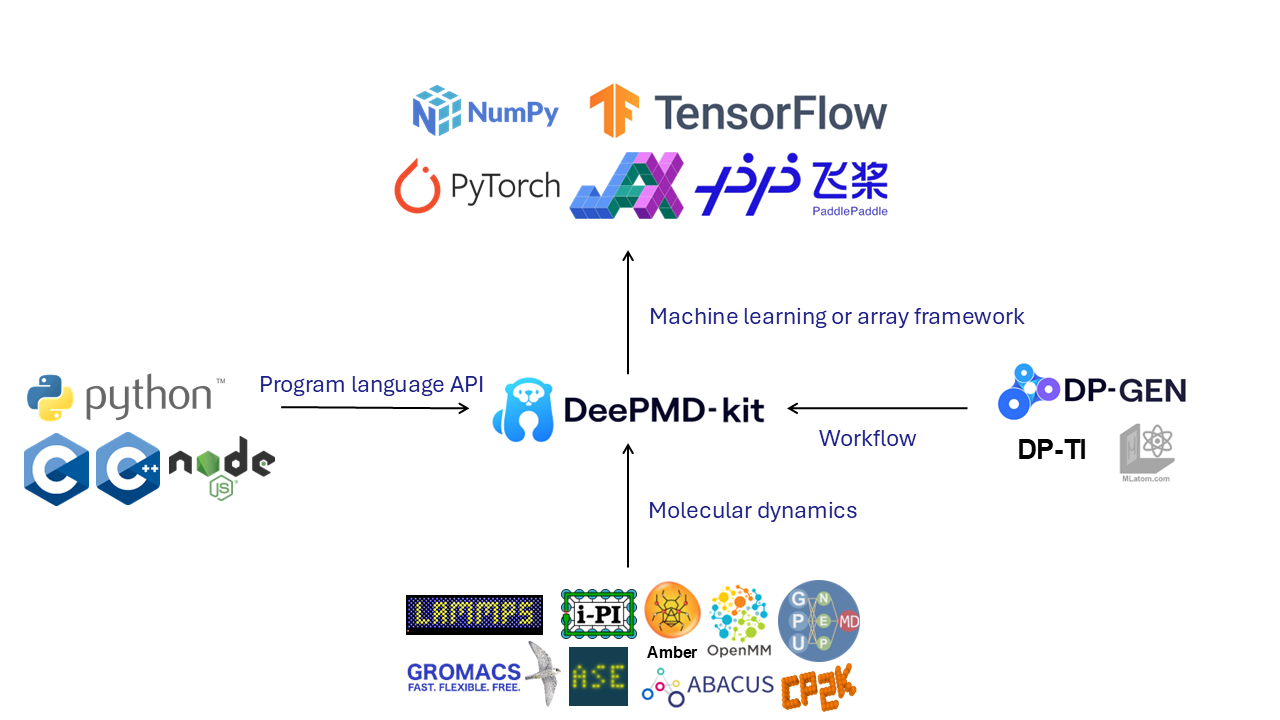}
    \centering
    \caption{The DeePMD-kit software ecosystem.
    The arrows indicate dependency flow.
    Software packages shown in the figure include:
    (1) DeePMD-kit;
    (2) Machine learning and array frameworks: NumPy\cite{Harris_Nature_2020_v585_p357}, TensorFlow\cite{Abadi_2015_tensorflow}, PyTorch\cite{Paszke_arXiv_2019_p1912.01703}, JAX\cite{jax2018github}, and PaddlePaddle\cite{Ma_FrontDataComput_2019_v1_p105};
    (3) Molecular dynamics packages: LAMMPS\cite{Thompson_ComputPhysCommun_2022_v271_p108171}, i-PI\cite{Litman_JChemPhys_2024_v161_p062504}, Amber\cite{Case_JChemInfModel_2023_v63_p6183}, OpenMM\cite{Eastman_JPhysChemB_2024_v128_p109}, CP2K\cite{Kuhne_JChemPhys_2020_v152_p194103}, GROMACS\cite{Abraham_SoftX_2015_v1_p19}, ASE\cite{HjorthLarsen_JPhysCondensMatter_2017_v29_p273002}, ABACUS\cite{Li_ComputationalMaterialsScience_2016_v112_p503}, and GPU-MD\cite{Fan_ComputPhysCommun_2017_v218_p10};
    (4) Workflow packages: DP-GEN\cite{Zhang_ComputPhysCommun_2020_v253_p107206}, MLatom\cite{Dral_JChemTheoryComput_2024_v20_p1193}, and DP-TI\cite{dpti};
    (5) Program language API: Python, C, C++, and Node.js.
    }
    \label{fig:ecosystem}
\end{figure}

In this work, we introduce a major new release of the DeePMD-kit package (v3),
featuring a multi-backend framework that integrates the existing TensorFlow\cite{Abadi_2015_tensorflow} backend with new backends, including PyTorch\cite{Paszke_arXiv_2019_p1912.01703}, JAX\cite{jax2018github}, and PaddlePaddle\cite{Ma_FrontDataComput_2019_v1_p105}.
These backends are designed to be interchangeable, allowing users and developers to switch between them with minimal modifications.
The new framework reuses the well-established interfaces developed in previous versions (see Fig.~\ref{fig:ecosystem}), maintaining seamless integration with other software, such as
LAMMPS\cite{Thompson_ComputPhysCommun_2022_v271_p108171}, i-PI\cite{Litman_JChemPhys_2024_v161_p062504}, AMBER\cite{Case_JChemInfModel_2023_v63_p6183,Tao_JChemPhys_2024_v160_p224104,Giese_JPhysChemB_2024_v128_p6257}, CP2K\cite{Kuhne_JChemPhys_2020_v152_p194103}, OpenMM\cite{Eastman_JPhysChemB_2024_v128_p109,Ding_IntJMolSci_2024_v25_p1448}, GROMACS\cite{Abraham_SoftX_2015_v1_p19}, ASE\cite{HjorthLarsen_JPhysCondensMatter_2017_v29_p273002}, and ABACUS\cite{Li_ComputationalMaterialsScience_2016_v112_p503}.
The inclusion of these new backends enables DeePMD-kit to seamlessly interact with other packages that utilize PyTorch, JAX, or PaddlePaddle, thereby expanding its compatibility and functionality.
Additionally, this update enables DeePMD-kit to leverage unique features and optimizations available in different machine learning frameworks, further enhancing its flexibility and performance.

\section{Software Description}

\subsection{A Multiple-backend Framework}

The multiple-backend framework added in DeePMD-kit v3 is aimed to support multiple deep-learning frameworks in a pluggable way
while providing a unified interface for users and developers.
When introducing the multi-backend framework, no breaking changes were made to the existing Python and C/C++ APIs in DeePMD-kit v2.
As a result, the existing interfaces implemented in various molecular dynamics packages can be reused without requiring any modifications.

\begin{figure}[t]
    \includegraphics[width=\linewidth]{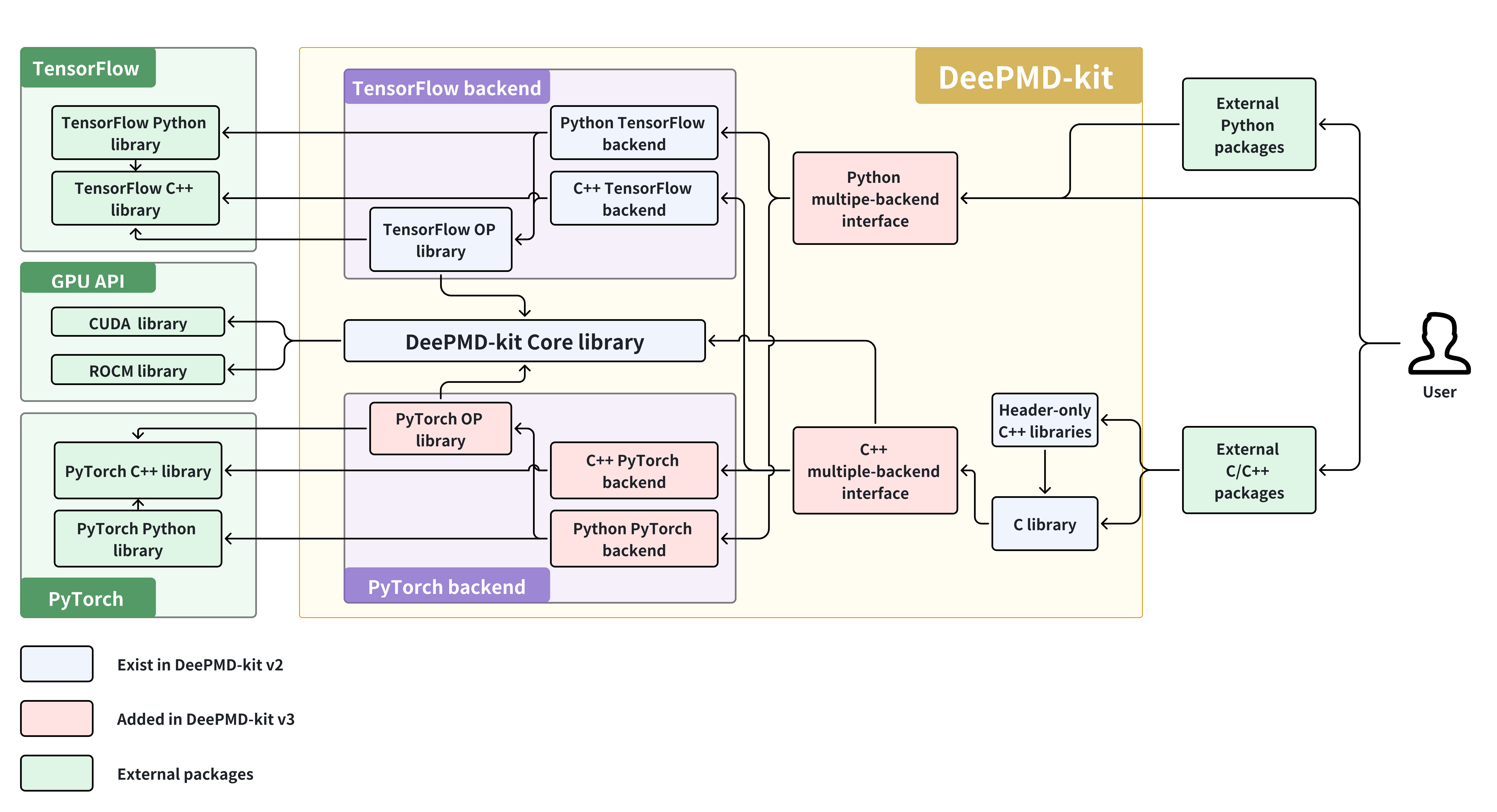}
    \centering
    \caption{The infrastructure of the TensorFlow and PyTorch backends in the DeePMD-kit software.
    The arrows indicate dependency flow.
    Modules in blue existed in DeePMD-kit v2\cite{Zeng_JChemPhys_2023_v159_p054801},
    modules in red are newly added in DeePMD-kit v3,
    and modules in green are machine learning frameworks.
    }
    \label{fig:framework}
\end{figure}

The infrastructure of the TensorFlow and PyTorch backends is illustrated in Fig.~\ref{fig:framework};
other backends are organized in a similar manner.
For both training and inference tasks driven by different models, users interact with a unified set of interfaces, regardless of the backend being used.
The backend-specific implementation is then invoked to perform the actual computations, leveraging external machine learning frameworks that are highly optimized for performance.
In some cases, these backends also share common implementations through customized operators.
For example, the model compression functionality\cite{Lu_JChemTheoryComput_2022_v18_p5559} is implemented in the DeePMD-kit core library, utilizing DeePMD-kit CUDA or ROCm libraries for efficient computation.

Users are expected to interact with each backend in a uniform manner,
without needing to understand the backend-specific details.
During model training and saving via the Python interface, users can simply specify the desired backend.
All other training parameters are designed to be backend-agnostic, ensuring a consistent user experience.
Once training is complete, the model is saved in a backend-specific format.
When the saved model is loaded for inference through the Python or C++ API during molecular dynamics simulations, the backend is automatically detected based on the model filename, and the appropriate backend module is used.
Furthermore, models created with one backend can be easily converted to another backend, providing flexibility and interoperability for various workflows.
An example user script is shown as follows:
\begin{lstlisting}[language=bash]
# Use the PyTorch backend to train and freeze the model
dp --pt train input.json
dp --pt freeze -o model.pth
# Test the model
dp test -m model.pth -s dataset
# Convert the model to that in the JAX backend
dp convert-backend model.pth model.savedmodel
\end{lstlisting}

Developers can incorporate new backends in a modular and pluggable manner,
focusing on backend-specific implementations without needing to modify the existing Python or C/C++ APIs or interfaces with external packages.
DeePMD-kit provides an implementation in the Array API\cite{Meurer_ProcPythonSciConf_2023_p8} (see Section~\ref{sec:backends}),
simplifying the process of adding new backends based on machine learning frameworks that support the Array API.
For developers looking to add new models, the Array API can also be used to implement these models for backends that support it.
For backends that do not support the Array API, model implementation can be facilitated with the assistance of large language models (LLMs), as shown in the Supporting Information, streamlining the development process further.

A challenge in the multiple-backend framework is to ensure that the same model driven by different backends produces the same results.
To address this issue, we have developed a set of tests that compare the results of the same model driven by different backends.
The models can be serialized and deserialized in the Python interface, which is the reason why they can be converted to each other.

\subsection{Backends}
\label{sec:backends}

Five backends are supported in the current version of DeePMD-kit: DP, TensorFlow\cite{Abadi_2015_tensorflow}, PyTorch\cite{Paszke_arXiv_2019_p1912.01703}, JAX\cite{jax2018github}, and PaddlePaddle\cite{Ma_FrontDataComput_2019_v1_p105}.

\paragraph{DP.}
The DP backend serves as a reference implementation, designed to provide a correct and standardized foundation for model development.
It is built using the Array API\cite{Meurer_ProcPythonSciConf_2023_p8}, allowing its functionality to be leveraged by other backends (such as JAX) without requiring code duplication.
By default, the DP backend uses NumPy\cite{Harris_Nature_2020_v585_p357}, which does not support gradient computations or GPU acceleration, for computations to minimize dependencies.

\paragraph{TensorFlow.}
The TensorFlow backend is the original backend of DeePMD-kit.
It utilizes the TensorFlow v1 API\cite{Abadi_2015_tensorflow}, which employs static computational graphs to optimize performance.
These static graphs can be saved into model files and later restored for inference in both Python and C++ interfaces.
The customized TensorFlow C++ operators are mainly used to calculate coordinate matrix, force, virial, and embedding network and matrix in compressed models.

\paragraph{PyTorch.}
The PyTorch backend leverages dynamic computational graphs to provide greater flexibility.\cite{Paszke_arXiv_2019_p1912.01703}
It employs TorchScript for model serialization, enabling models to be saved and loaded in both Python and C++ interfaces.
The customized PyTorch C++ operators are used for model compression and communication between processors or GPU cards in the graph neural networks of the DPA-2 model\cite{Zhang_npjComputMater_2024_v10_p293}.
PyTorch is friendly to be developed with and thus has a larger user base.
PyTorch is widely used in various atomistic packages, including machine learning potential packages\cite{Gao_JChemInfModel_2020_v60_p3408,Batatia_BookChap_NIPS_2022_v36_p830,Batzner_NatCommun_2022_v13_p2453}, force field packages\cite{Orlando_BioinformOxfordEngl_2024_v40_pbtae160}, semiempirical quantum chemistry packages\cite{Friede_JChemPhys_2024_v161_p062501}, and molecular dynamics packages\cite{Doerr_JChemTheoryComput_2021_v17_p2355}.
The development of the PyTorch backend has made it easier to integrate DeePMD-kit with these existing packages.

\paragraph{JAX.}
The JAX backend is built on top of the DP backend and the Array API\cite{Meurer_ProcPythonSciConf_2023_p8}, using JAX\cite{jax2018github} as its array library.
JAX performs just-in-time (JIT) compilation for enhanced performance.
Models created with the JAX backend are saved in a TensorFlow-compatible format using the ``jax2tf'' converter, allowing them to be loaded via the TensorFlow C++ library \cite{Abadi_2015_tensorflow}.
JAX has gained popularity in the field of atomistic simulations, supporting machine learning potential packages\cite{Orlando_BioinformOxfordEngl_2024_v40_pbtae160}, machine learning density functional packages\cite{Chen_ComputerPhysicsCommunications_2023_v282_p108520}, force field packages\cite{Wang_JChemTheoryComput_2023_v19_p5897,Kaymak_JChemTheoryComput_2022_v18_p5181}, and molecular dynamics packages\cite{Schoenholz_JStatMech_2021_v2021_p124016}.

A challenge in the JAX backend is that the JAX JIT compiler requires the input tensors to have static shapes, a constraint not imposed by TensorFlow or PyTorch.
However, during molecular dynamics simulations, parameters such as the maximum number of neighbor atoms, the simulation box size, and the number of ghost atoms that do not directly contribute atomic energies are dynamic and not known in advance.
To address this limitation, the compiled JIT model is wrapped within a standard TensorFlow model.
During each simulation step, the TensorFlow model calculates or post-processes the neighbor list and other tensors with dynamic shapes.
The neighbor list is then passed into the compiled JIT model, combining the strengths of both TensorFlow (for handling dynamic inputs) and JAX (for optimized performance through JIT compilation).
To handle the dynamic number of ghost atoms, a larger fixed value is initially set, which is adjusted if it becomes insufficient during the simulation.

\paragraph{PaddlePaddle.}
The PaddlePaddle backend, which is based on PaddlePaddle\cite{Ma_FrontDataComput_2019_v1_p105}, features a similar Python interface to the PyTorch backend, ensuring compatibility and flexibility in model development.
PaddlePaddle has introduced dynamic-to-static functionality and PaddlePaddle JIT compiler (CINN) in DeePMD-kit,
which allow for dynamic shapes and higher-order differentiation.
The dynamic-to-static functionality automatically captures the user's dynamic graph code and converts it into a static graph.
After conversion, the CINN compiler is used to optimize the computational graph,
thereby enhancing the efficiency of model training and inference.
In our experiments with the DPA-2 model,
we achieved approximately a 40\% reduction in training time compared to the dynamic graph,
effectively improving the model training efficiency.

\subsection{New Design Principles in version 3}

When new backends, including DP, PyTorch, JAX, and PaddlePaddle, were introduced, the design of the DeePMD-kit was thoroughly reconsidered.
The following design principles were adopted, differing significantly from those used in the previous TensorFlow backend of DeePMD-kit version 2:

\paragraph{Metaprogramming.}
To reduce code duplication and enhance maintainability, metaprogramming techniques are employed to generate backend-specific classes from those in the reference DP backend.
For instance, in different backends, a neural network is constructed from multiple neural network layers. While the structure of these layers is consistent across backends, the implementation must interact with different machine learning frameworks depending on the backend.
By transforming a base neural network class into a backend-specific neural network class using a backend-specific layer class, the implementation of neural networks becomes backend-agnostic.
This approach ensures a consistent structure while allowing flexibility for backend-specific details, significantly streamlining development and maintenance.

\paragraph{Atomic model.}

\begin{figure}[t]
  \includegraphics[width=\linewidth]{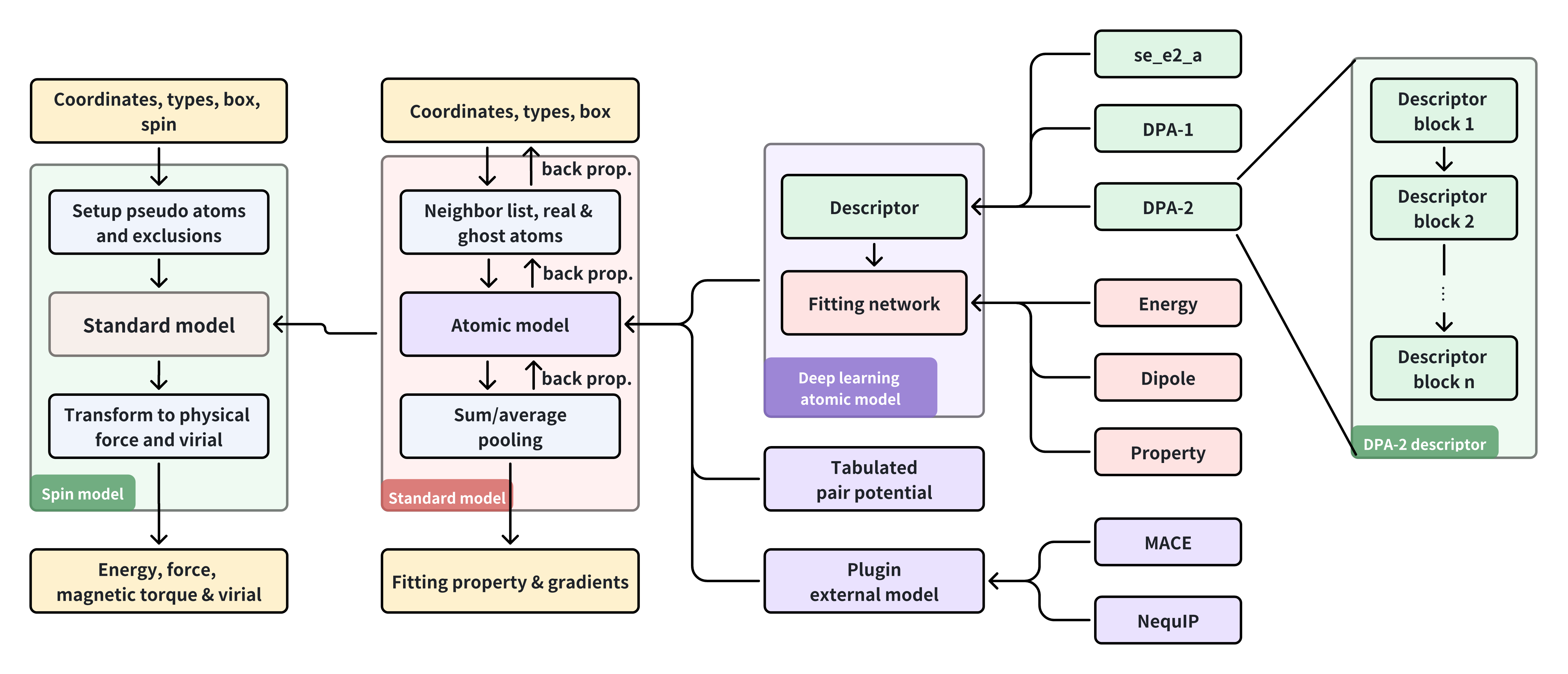}
  \centering
  \caption{Schematic plot of the model components in DeePMD-kit version 3.}
  \label{fig:model}
\end{figure}

In version 3, we introduce an innovative component termed the ``atomic model'' (see Fig.~\ref{fig:model}). This model is based on the assumption that the physical quantity to be learned, represented by \(\bm y\), can be decomposed into atomic contributions, expressed as:
\begin{align}
  \bm y = \sum_i \bm y_i
\end{align}
This approach simplifies the implementation process for model developers, who only need to define and implement the atomic contributions \(\bm y_i\).
The DeePMD-kit subsequently performs the summation of these atomic contributions and computes the generalized force $\bm{F}_i$ and virial $\Xi_{\alpha, \beta}$,
which are defined as the partial derivatives of the physical quantity with respect to atomic coordinates and cell vectors, respectively:
\begin{align}\label{eq:fv}
  & \bm F_i =  -\frac{\partial \boldsymbol y}{\partial \boldsymbol r_i}, \qquad
    \bm \Xi_{\alpha\beta} =   -\sum_{\gamma} \frac{\partial \boldsymbol y}{\partial h_{\gamma\alpha}} h_{\gamma\beta}.
\end{align}
In Equation \eqref{eq:fv}, \(\boldsymbol r_i\) represents the Cartesian coordinates of atom \(i\), while \(h_{\alpha\beta}\) denotes the \(\beta\) component of the \(\alpha\)-th cell vector.

\paragraph{Computation of the neighbor list and the coordinate matrix.}
In the TensorFlow backend, the neighbor list, the coordinate matrix, and the gradient of the coordinate matrix with respect to the coordinates were computed using custom TensorFlow operators,
as these operations are not standard in TensorFlow\cite{Wang_ComputPhysCommun_2018_v228_p178}.
This design posed challenges for calculating and training the Hessian, the second gradient of the output with respect to atomic coordinates,
$\frac{\partial \boldsymbol{y}}{\partial \boldsymbol{x}_i \partial \boldsymbol{x}_j}$.
In the new backends, these computations are performed using standard operators provided by the respective machine learning frameworks.
This enables straightforward inference and training of the Hessian using the automatic differentiation capabilities inherent to these frameworks.
The neighbor list can be configured to exclude specific pairs of atomic types, thereby implementing range corrections for MD simulations using MLP/MM or QM/MM+$\Delta$MLP\cite{Zeng_JChemTheoryComput_2021_v17_p6993}.

\paragraph{Implementation of communications for graph neural network models.}
The new backends now support graph neural network (GNN) models, such as the DPA-2 model\cite{Zhang_npjComputMater_2024_v10_p293} and the in-development DPA-3 model.
A key challenge in implementing GNN models is managing communication between nodes or GPU cards during molecular dynamics simulations.
In a GNN model, atomic features are updated based on atom and edge features within a cutoff radius, which requires access to the neighbor list of surrounding atoms.
However, some neighbor atoms may be ghost atoms that do not directly contribute atomic energies, and their neighbor lists are typically unavailable in most molecular dynamics packages.\cite{Thompson_ComputPhysCommun_2022_v271_p108171}
Some GNN-based MLP packages address this issue by generating ghost atoms within an extended cutoff radius ($r_c \times L$, where $r_c$ is the cutoff radius for each layer and $L$ is the number of layers) and rebuilding the neighbor lists for the ghost atoms.
This approach, however, significantly increases computational cost.
Other packages have suggested running simulations on a single GPU card to avoid this issue.
To overcome these limitations, the new backends implement a customized C++ operator that uses the message-passing interface (MPI)\cite{mpi41} to exchange atom and edge features between processors or GPU cards.
This solution enables efficient simulations across multiple nodes or GPU cards in high-performance computing environments.

\paragraph{Descriptor block.}
The complexity of descriptors has increased significantly with the advancement of deep potential models.
The DPA-2 model\cite{Zhang_npjComputMater_2024_v10_p293}, for instance, primarily consists of a representation-initializer block,
representation-transformer layers, and a three-body embedding block.
The representation-initializer block is derived from the DPA-1 descriptor\cite{Zhang_npjComputMater_2024_v10_p94}.
To address the growing complexity and support future development,
we introduce a novel component called the ``descriptor block'' (see Fig.~\ref{fig:model}), which serves as a modular foundation for constructing complex descriptors.
The descriptor block is designed to be flexible and extensible, allowing seamless integration of new descriptors into the framework.

\paragraph{Refactor of the implementation of the DeepSPIN model.}
In the TensorFlow backend, the DeepSPIN model~\cite{Yang_PhysRevB_2024_v110_p64427} was initially integrated into the descriptors and fitting networks in an ad hoc manner, which hindered maintainability and limited the adoption of new features such as the DPA-2 descriptor.
To address these limitations, the DeepSPIN model has been refactored in the new backends into a modular component that operates on top of any standard potential energy model.

The DeepSPIN model aims to represent the potential energy of a system as a function of atomic coordinates, spin, and the simulation cell.
To achieve this, the model introduces pseudo atoms, which are associated with physical atoms through the relation:
\begin{align}
    \bm r_{i^p} = \bm r_i + \lambda_{\eta_i}\bm{S}_i, \quad i\in \mathcal A,
\end{align}
where \(\mathcal A\) denotes the set of physical atoms, \(\bm{S}_i\) is the spin of atom \(i\), \(\lambda_{\eta_i}\) is a user-defined, element-specific scaling factor (with \(\eta_i\) representing the element type of atom \(i\)), and \(\bm r_i\) is the Cartesian coordinate of physical atom \(i\).
Notably, pseudo atoms do not represent physical entities but serve as computational tools to encode the spin states of their associated physical atoms.
To ensure type consistency, pseudo atoms inherit mapped element types from their host atoms (e.g., \texttt{Fe\_spin} from \texttt{Fe}), denoted by \(\eta_{i^p}^\text{spin}\).

The model processes an augmented atomic input, which includes both the augmented coordinates \(\{\bm r_i, \bm r_{i^p} \,|\, i \in \mathcal{A}\}\) and the augmented types \(\{\eta_i, \eta_{i^p}^\text{spin} \,|\, i \in \mathcal{A}\}\).
These inputs are fed into a standard potential energy model to generate augmented outputs, including the augmented atomic energies \(\{\mathcal{E}_i, \mathcal{E}_{i^p} \,|\, i \in \mathcal{A}\}\) and the augmented atomic forces \(\{\hat{\bm F}_i, \hat{\bm F}_{i^p} \,|\, i \in \mathcal{A}\}\).
Post-processing steps are then applied to convert the augmented predictions into physical quantities. The total energy is computed as the sum of energy contributions from all physical atoms:
\begin{align}
   E = \sum_{i \in \mathcal{A}} E_i.
\end{align}
The atomic forces \(\bm F_i\) and magnetic torques \(\bm{\omega}_i\) for atom \(i\) are derived, respectively, as:
\begin{align}
   \bm F_i = -\frac{\partial E}{\partial \bm r_i} = \hat{\bm F}_i + \hat{\bm F}_{i^p}, \quad \bm{\omega}_i = -\frac{\partial E}{\partial \bm{S}_i} = \lambda_{\eta_i} \hat{\bm F}_{i^p}.
\end{align}
The system virial \(\bm{\Xi}\) is calculated by summing the outer products of the forces and positions of all physical atoms~\cite{Tuckerman_Book_StatMech_2010}:
\begin{align}
   \bm{\Xi} = \sum_{i \in \mathcal{A}} \bm F_i \otimes \bm r_i.
\end{align}
This refactored design, as illstrated by Fig.~\ref{fig:model}, enhances code clarity and maintainability while maintaining compatibility with standard potential energy model architectures.

\paragraph{Removal of the local frame descriptor.}
In the previous publication\cite{Zeng_JChemPhys_2023_v159_p054801}, we have indicated that the local frame descriptor\cite{Zhang_PhysRevLett_2018_v120_p143001} is not continous
at the cutoff radius and the exchanging of the order of two nearest neighbors.
Thus, the local frame descriptor is removed in the new backends.

\section{Selected Extensions}

DeePMD-kit v3 introduces an extensive plugin system designed to seamlessly integrate external models into its various backends.
In this section, we demonstrate the versatility of the multi-backend framework by showcasing the integration of other MLPs packages and of Differentiable Molecular Force Field.

\subsection{DeePMD-GNN}
Most machine learning potential (MLP) packages, including the previous version of DeePMD-kit, are typically restricted to a single type of neural network potential developed by the same team of developers.
This limitation imposes additional burdens on users, who must learn new packages, and on developers, who must maintain multiple packages, while also makes it impossible to benchmark models using the same loss function, learning rate, and data sets.
In the earlier version of DeePMD-kit, integrating other MLP packages was challenging because most were not based on TensorFlow.
This limitation was addressed with the development of the multi-backend framework and the PyTorch backend.
Consequently, Zeng \textit{et al.} introduced the DeePMD-GNN package\cite{DEEPMD_GNN}, which integrates the MACE\cite{Batatia_BookChap_NIPS_2022_v36_p830} and NequIP\cite{Batzner_NatCommun_2022_v13_p2453} models into the PyTorch backend.
This advancement marked the first time DeePMD-kit supported highly mature external potential models, significantly enhancing its versatility.
It is worth noting that the plugin solely extends the trainable model module (see Fig.~\ref{fig:model}) and reuses the existing loss function and learning rate modules in the DeePMD-kit, allowing different models to be compared easily under the same conditions.

\subsection{DMFF plugin}
Long-range interactions are critical in many systems but are challenging to incorporate into an MLP model\cite{Anstine_JPhysChemA_2023_v127_p2417}.
Zhu \textit{et al.}\cite{Zhu_DMFF_plugin_for_2025} developed a PyTorch version of the Differentiable Molecular Force Field (DMFF) package\cite{Wang_JChemTheoryComput_2023_v19_p5897} that implements Ewald summation\cite{Darden_JChemPhys_1993_v98_p10089} and charge equilibration (QEq)\cite{Rappe_JPhysChem_1991_v95_p3358} methods for handling long-range interactions.
These interactions have been integrated into the PyTorch backend of DeePMD-kit, enabling a hybrid approach where long-range interactions are described by traditional force fields, while short- and mid-range interactions are captured by MLPs.

\section{Benchmark}

A key application of the multi-backend framework is its ability to run molecular dynamics simulations using the most efficient backend.
To illustrate this, we present benchmark calculations using the following models:
DPA-1 without attention layers (L=0) and its compressed version\cite{Lu_JChemTheoryComput_2022_v18_p5559},
DPA-1 with two attention layers (L=2)\cite{Zhang_npjComputMater_2024_v10_p94},
and DPA-2 (medium)\cite{Zhang_npjComputMater_2024_v10_p293}.
These benchmarks were performed in both single- and double-precision modes, employing the TensorFlow, PyTorch, and JAX backends on a single 80~GB NVIDIA H100 GPU card, 40~GB NVIDIA A800 GPU card, and 24~GB NVIDIA 4090 GPU card.
The simulations involved a water system of varying atom counts, with each calculation repeated 500 times to obtain an average computational speed.
LAMMPS\cite{Thompson_ComputPhysCommun_2022_v271_p108171} interfaced with DeePMD-kit v3 is used to perform simulations.
Here we note the limitations in the current package when this article was written:
only the DPA-1 (L=0) model supports model compression,
the TensorFlow backend does not support the DPA-2 model,
the JAX backend does not support model compression,
and the PaddlePaddle backend is still under development.

All of the models use a fitting network that consists of 3 hidden layers with 240, 240, and 240 neurons.
The coordinate matrix is encoded from the local environment within a 6 \AA{}  cutoff radius and 1 \AA{} of smoothing;
The details of DPA-1 models are as below:
the embedding network consists of 3 hidden layers with 25, 50, and 100 neurons;
the embedding submatrix size is 12.
Specifically, in the DPA-1 (L=2) model, the length of hidden vectors in each attention layer is 128.
The details of the DPA-2 (medium) model are as below:
the representation initializer layer is encoded based on the local environment within a 6 \AA{} cutoff radius and smoothed over 1 \AA{};
the model uses 6 representation transformer (Reperformer) layers, with each layer calculated using a 4 \AA{} cutoff and 1 \AA{} of smoothing;
three-body embedding is incorporated within a 4 \AA{} cutoff;
the embedding network consists of three hidden layers with 25, 50, and 100 neurons, and the embedding submatrix has a size of 12;
the fitting network also has three hidden layers, each containing 240 neurons;
the dimensions for the invariant single-atom and pair-atom representations are set to 120 and 32, respectively.
Additionally, the localized single-atom representation update mechanism excludes the self-attention layer, while the Reperformer pair-atom representation is updated using a gated self-attention layer.

\begin{figure}[t]
  \includegraphics[width=\linewidth]{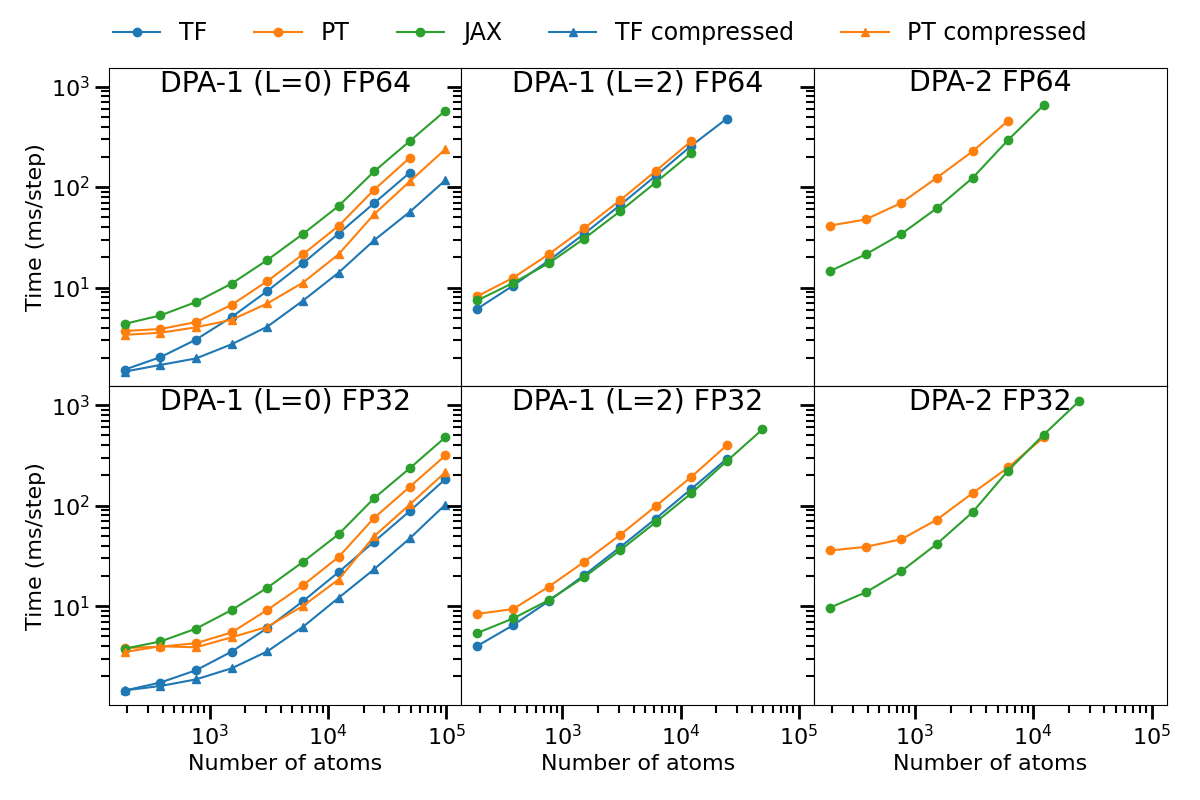}
  \centering
  \caption{The molecular dynamics performance (ms/step) of DPA-1 (L=0), DPA-1 (L=2), and DPA-2 models in double (FP64) and single (FP32) precisions in TensorFlow (TF), PyTorch (PT), and JAX backends on a single 80~GB NVIDIA H100 GPU card.
  The water system with different numbers of atoms is used for simulations.
  }
  \label{fig:h100}
\end{figure}

The performance on the H100, A800, and 4090 cards is summarized in Table~S1, S2, and S3, respectively,
and illustrated in Fig.~\ref{fig:h100}, S1, and S2, respectively.
Across TensorFlow, PyTorch, and JAX, the three GPU cards exhibit similar relative performance.
The results indicate that for the DPA-1 model without attention layers, the compressed TensorFlow model is the fastest.
Although model compression improves performance by a similar amount on both the TensorFlow and PyTorch backends, other components of the model cause the final performance to differ.
For more computationally demanding models, such as the DPA-1 model with attention layers and the DPA-2 model, the JAX backend generally achieves the highest performance.
However, two exceptions were observed: in the DPA-2 FP32 case with 12,288 atoms on the H100 card, where PyTorch outperformed JAX; in the DPA-1 (L=2) FP64 case on the 4090 card, where TensorFlow outperformed JAX.
Additionally, different backends exhibit varying GPU memory usage, which can influence the choice of backend depending on the computational resources available.
These findings demonstrate that no single backend consistently outperforms others across all models.
Therefore, the multi-backend framework is highly valuable, enabling users to identify and utilize the most efficient backend for their specific needs.

\section{Conclusions}
In this work, we introduced DeePMD-kit v3, a significant enhancement to the widely used machine learning potential package.
This new version addresses critical limitations of its predecessor by incorporating a multi-backend framework supporting TensorFlow, PyTorch, JAX, and PaddlePaddle.
By enabling seamless backend switching with minimal changes, DeePMD-kit v3 provides a flexible and interoperable platform for researchers and developers.
This advancement not only facilitates the integration of DeePMD-kit with diverse software ecosystems but also empowers the development of more complex and efficient workflows for molecular dynamics simulations and other applications requiring machine learning potentials.
The best performance for molecular dynamics can be archeived by selecting the most efficient backend for a specific model and computational resource.
We anticipate that DeePMD-kit v3 will significantly expand the accessibility and applicability of MLPs in physics, chemistry, and materials science, fostering innovation and collaboration in these fields.

\section*{Data Availability}

Source code for the project can be found at \url{https://github.com/deepmodeling/deepmd-kit}.
Source code for the benchmark can be found at \url{https://github.com/njzjz/benchmark-dpv3}.

\begin{acknowledgement}

The authors thank Dr.~Rocco Meli for his code contributions.
ChatGPT 4o was used to edit the English (prompt: polish the following paragraph in a research article).
J. Z. acknowledges 2023 Chinese Government Award for Outstanding Self-financed Students Aboard presented by the China Scholarship Council.
S. H. acknowledges help from Yanjun Ma, Tiezhu Gao, Xiaoguang Hu.
J.-X. Z. gratefully acknowledges Xiamen University and iChEM for a Ph.D. studentship.
A. K. S. acknowledges help from Dr.~Manish Modani (Principal Solution Architect at NVIDIA), Dr. Unnikrishnan C (Assistant Professor at IIT, Palakkad), and C-DAC Centre, Patna, India.
The work of Q. Z. and J. D. was supported by the the fundation the Science and Technology Innovation Program of Hunan Province under Grant No.~2021RC4026.
The work of S. L. B. was supported by the Research Council of Norway through the Centre of Excellence Hylleraas Centre for Quantum Molecular Sciences (Grant 262695) and the Young Researcher Talent (Grant 344993), as well as by the EuroHPC Joint Undertaking (Grant EHPC-REG-2023R02-088).
The work of W. J. was supported by the Natural Science foundation of China (No. 92270206).
The work of T. J. G. and D. M. Y. was supported by the National Institutes of Health (No. GM107485 to D.M.Y.)
and
the National Science Foundation (CSSI Frameworks Grant No. 2209718 to D.M.Y).
The work of H. W. was supported by the National Key R\&D Program of China (Grant No.~2022YFA1004300) and the National Natural Science Foundation of China (Grant No.~12122103).
Computational resources are provided by:
ByteDance Volcano Engine Cloud;
Alibaba Cloud PAI;
IKKEM Intelligent Computing Center, for the work of Y.-P. L.

\end{acknowledgement}

\begin{suppinfo}

The molecular dynamics performance (ms/step) of DPA-1 (L=0), DPA-1 (L=2), and DPA-2 models in double (FP64) and single (FP32) precisions in TensorFlow (TF), PyTorch (PT), and JAX backends on a single 40~GB NVIDIA H100, A800, and 4090 GPU card.
Example of Using Large Language Models for Code Conversion between Backends.

\end{suppinfo}

\bibliography{combined,tmplib}

\end{document}


\section{Tables and Figures}

\begin{table}[]
    \resizebox{\textwidth}{!}{
    \begin{tabular}{l|r|rrrrr|rrr|rr}
    \multirow{2}{*}{Precision} & \multirow{2}{*}{\# atoms} & \multicolumn{5}{c|}{DPA-1 (L=0)}                                              & \multicolumn{3}{c|}{DPA-1 (L=2)}                                             & \multicolumn{2}{c}{DPA-2}                         \\
                               &                          & TF                      & PT                      & JAX    & TFc    & PTc    & TF                      & PT                      & JAX                     & PT                      & JAX                     \\
    \hline
    \multirow{10}{*}{FP64}     & 192                                           & 1.53                    & 3.71                    & 4.36   & \textbf{1.46}   & 3.40   & \textbf{6.18}                    & 8.23                    & 7.47                    & 41.46                   & \textbf{14.62}                   \\
                               & 384                                           & 2.04                    & 3.88                    & 5.29   & \textbf{1.70}   & 3.56   & \textbf{10.49}                   & 12.52                   & 11.15                   & 47.62                   & \textbf{21.42}                   \\
                               & 768                                           & 3.04                    & 4.55                    & 7.20   & \textbf{1.98}   & 4.03   & 18.56                   & 21.48                   & \textbf{17.47}                   & 69.71                   & \textbf{34.18}                   \\
                               & 1536                                          & 5.11                    & 6.75                    & 10.98  & \textbf{2.74}   & 4.79   & 34.37                   & 39.05                   & \textbf{30.80}                   & 124.00                  & \textbf{61.77}                   \\
                               & 3072                                          & 9.26                    & 11.60                   & 18.83  & \textbf{4.09}   & 6.96   & 66.00                   & 73.82                   & \textbf{57.35}                   & 226.73                  & \textbf{123.22}                  \\
                               & 6144                                          & 17.50                   & 21.47                   & 34.19  & \textbf{7.41}   & 11.22  & 129.22                  & 144.01                  & \textbf{110.95}                  & 457.75                  & \textbf{294.68}                  \\
                               & 12288                                         & 34.46                   & 41.09                   & 64.76  & \textbf{14.09}  & 21.46  & 257.11                  & 286.22                  & \textbf{217.21}                  & OOM & \textbf{655.23}                  \\
                               & 24576                                         & 69.32                   & 94.21                   & 143.25 & \textbf{29.62}  & 53.78  & \textbf{479.70}                  & OOM & OOM & OOM & OOM \\
                               & 49152                                         & 139.24                  & 195.81                  & 286.93 & \textbf{56.72}  & 113.98 & OOM & OOM & OOM & OOM & OOM \\
                               & 98304                                         & OOM & OOM & 575.62 & \textbf{117.23} & 238.53 & OOM & OOM & OOM & OOM & OOM \\
    \hline
    \multirow{10}{*}{FP32}     & 192                                           & \textbf{1.45}                    & 3.84                    & 3.78   & 1.46   & 3.50   & \textbf{4.04}                    & 8.39                    & 5.42                    & 35.83                   & \textbf{9.69}                    \\
                               & 384                                           & 1.74                    & 3.96                    & 4.46   & \textbf{1.61}   & 4.00   & \textbf{6.47}                    & 9.37                    & 7.56                    & 38.94                   & \textbf{13.74}                   \\
                               & 768                                           & 2.31                    & 4.29                    & 5.99   & \textbf{1.87}   & 3.90   & \textbf{11.22}                   & 15.62                   & 11.49                   & 46.35                   & \textbf{22.28}                   \\
                               & 1536                                          & 3.54                    & 5.49                    & 9.17   & \textbf{2.41}   & 4.91   & 20.40                   & 27.66                   & \textbf{19.53}                   & 72.60                   & \textbf{41.72}                   \\
                               & 3072                                          & 6.07                    & 9.20                    & 15.27  & \textbf{3.56}   & 6.24   & 38.50                   & 50.98                   & \textbf{35.85}                   & 133.20                  & \textbf{86.00}                   \\
                               & 6144                                          & 11.21                   & 16.14                   & 27.57  & \textbf{6.23}   & 10.05  & 74.11                   & 98.62                   & \textbf{68.26}                   & 239.83                  & \textbf{223.71}                  \\
                               & 12288                                         & 21.82                   & 30.84                   & 51.97  & \textbf{12.11}  & 18.43  & 146.37                  & 192.70                  & \textbf{132.45}                  & \textbf{479.85}                  & 509.55                  \\
                               & 24576                                         & 43.92                   & 75.62                   & 118.79 & \textbf{23.34}  & 49.69  & 292.36                  & 398.49                  & \textbf{277.51}                  & OOM                    & \textbf{1,101.69}                \\
                               & 49152                                         & 88.37                   & 154.09                  & 235.74 & \textbf{47.29}  & 102.83 & OOM                    & OOM                    & \textbf{573.12}                  & OOM                    & OOM                    \\
                               & 98304                                         & 184.06                  & 315.71                  & 479.91 & \textbf{102.68} & 215.32 & OOM                    & OOM                    & OOM                    & OOM                    & OOM

    \end{tabular}
    }
    \caption{The molecular dynamics performance (ms/step) of DPA-1 (L=0), DPA-1 (L=2), and DPA-2 models in double (FP64) and single (FP32) precisions in TensorFlow (TF), PyTorch (PT), and JAX backends on a single 80~GB NVIDIA H100 GPU card.
    The water system with different numbers of atoms is used for simulations.
    ``TFc'' and ``PTc'' donate to compressed models in TensorFlow and PyTorch backends.
    ``OOM'' donates to out-of-memory (GPU memory exceeds 80~GB).
    The best performance for each model is highlighted in bold.
    }
    \label{tab:benchmark}
\end{table}

\begin{table}[]
    \resizebox{\textwidth}{!}{
    \begin{tabular}{l|r|rrrrr|rrr|rr}
    \multirow{2}{*}{Precision} & \multirow{2}{*}{\# atoms} & \multicolumn{5}{c|}{DPA-1 (L=0)}                                              & \multicolumn{3}{c|}{DPA-1 (L=2)}                                             & \multicolumn{2}{c}{DPA-2}                         \\
                               &                          & TF                      & PT                      & JAX    & TFc    & PTc    & TF                      & PT                      & JAX                     & PT                      & JAX                     \\
    \hline
    \multirow{10}{*}{FP64}
     & 192 & 2.10 & 3.89 & 5.55 & \textbf{1.62} & 3.66 & \textbf{10.71} & 13.19 & 11.77 & 46.04 & \textbf{22.37}  \\
     & 384 & 3.14 & 4.39 & 6.51 & \textbf{1.89} & 3.88 & 19.86 & 24.01 & \textbf{19.27} & 58.97 & \textbf{36.54}  \\
     & 768 & 5.22 & 6.63 & 9.38 & \textbf{2.55} & 4.53 & 36.63 & 43.88 & \textbf{33.45} & 97.69 & \textbf{64.70}  \\
     & 1536 & 9.42 & 11.42 & 15.40 & \textbf{3.92} & 6.02 & 70.10 & 82.85 & \textbf{61.55} & 185.50 & \textbf{123.88}  \\
     & 3072 & 17.85 & 21.56 & 28.15 & \textbf{6.85} & 9.68 & 136.80 & 160.26 & \textbf{117.91} & 365.57 & \textbf{239.78}  \\
     & 6144 & 35.05 & 40.74 & 53.93 & \textbf{12.99} & 18.38 & 270.16 & 315.86 & \textbf{231.34} & OOM & \textbf{537.82}  \\
     & 12288 & 69.48 & 80.32 & 103.68 & \textbf{25.91} & 36.13 & OOM & OOM & OOM & OOM & OOM  \\
     & 24576 & 138.06 & 180.34 & 236.52 & \textbf{51.05} & 83.99 & OOM & OOM & OOM & OOM & OOM  \\
     & 49152 & OOM & OOM & 473.12 & \textbf{101.63} & 197.73 & OOM & OOM & OOM & OOM & OOM  \\
     & 98304 & OOM & OOM & OOM & OOM & OOM & OOM & OOM & OOM & OOM & OOM  \\
    \hline
    \multirow{10}{*}{FP32}
     & 192 & 1.60 & 3.86 & 4.68 & \textbf{1.55} & 3.65 & \textbf{5.66} & 9.13 & 6.86 & 37.25 & \textbf{12.56}  \\
     & 384 & 2.13 & 4.06 & 5.49 & \textbf{1.74} & 3.80 & \textbf{10.26} & 15.40 & 10.53 & 41.60 & \textbf{19.15}  \\
     & 768 & 3.32 & 5.12 & 7.37 & \textbf{2.17} & 4.11 & 19.31 & 27.90 & \textbf{17.91} & 60.16 & \textbf{33.39}  \\
     & 1536 & 5.64 & 8.18 & 11.67 & \textbf{3.10} & 5.30 & 36.05 & 51.29 & \textbf{31.87} & 107.89 & \textbf{65.21}  \\
     & 3072 & 10.39 & 14.72 & 20.17 & \textbf{5.14} & 7.65 & 69.64 & 98.06 & \textbf{59.74} & 200.71 & \textbf{140.32}  \\
     & 6144 & 20.14 & 30.45 & 38.42 & \textbf{9.33} & 16.40 & 137.09 & 193.79 & \textbf{117.12} & 399.64 & \textbf{335.01}  \\
     & 12288 & 40.19 & 56.43 & 90.43 & \textbf{18.39} & 28.04 & 272.81 & 380.96 & \textbf{240.44} & OOM & \textbf{751.13}  \\
     & 24576 & 80.75 & 126.35 & 182.56 & \textbf{37.17} & 84.38 & OOM & OOM & \textbf{516.67} & OOM & OOM  \\
     & 49152 & 174.72 & 287.64 & 359.39 & \textbf{74.89} & 174.81 & OOM & OOM & OOM & OOM & OOM  \\
     & 98304 & OOM & OOM & 748.81 & \textbf{190.05} & 343.07 & OOM & OOM & OOM & OOM & OOM  \\
    \end{tabular}
    }
    \caption{The molecular dynamics performance (ms/step) of DPA-1 (L=0), DPA-1 (L=2), and DPA-2 models in double (FP64) and single (FP32) precisions in TensorFlow (TF), PyTorch (PT), and JAX backends on a single 40~GB NVIDIA A100 GPU card.
    The water system with different numbers of atoms is used for simulations.
    ``TFc'' and ``PTc'' donate to compressed models in TensorFlow and PyTorch backends.
    ``OOM'' donates to out-of-memory (GPU memory exceeds 40~GB).
    The best performance for each model is highlighted in bold.
    }
    \label{tab:benchmark}
\end{table}

\begin{table}[]
    \resizebox{\textwidth}{!}{
    \begin{tabular}{l|r|rrrrr|rrr|rr}
    \multirow{2}{*}{Precision} & \multirow{2}{*}{\# atoms} & \multicolumn{5}{c|}{DPA-1 (L=0)}                                              & \multicolumn{3}{c|}{DPA-1 (L=2)}                                             & \multicolumn{2}{c}{DPA-2}                         \\
                               &                          & TF                      & PT                      & JAX    & TFc    & PTc    & TF                      & PT                      & JAX                     & PT                      & JAX                     \\
    \hline
    \multirow{10}{*}{FP64}
     & 192 & 3.87 & 5.73 & 8.31 & \textbf{2.53} & 5.10 & \textbf{29.53} & 39.45 & 40.57 & 67.58 & \textbf{46.00}  \\
     & 384 & 6.65 & 8.97 & 12.76 & \textbf{3.08} & 5.65 & \textbf{60.22} & 76.42 & 76.20 & 101.39 & \textbf{78.15}  \\
     & 768 & 12.75 & 16.60 & 21.17 & \textbf{5.34} & 7.62 & \textbf{122.29} & 156.98 & 148.04 & 181.20 & \textbf{154.35}  \\
     & 1536 & 24.86 & 31.93 & 38.53 & \textbf{9.50} & 13.00 & \textbf{242.76} & 309.81 & 288.89 & 364.03 & \textbf{302.31}  \\
     & 3072 & 47.67 & 60.74 & 71.99 & \textbf{16.89} & 21.76 & \textbf{479.53} & 617.38 & 570.34 & OOM & \textbf{595.34}  \\
     & 6144 & 95.05 & 119.03 & 141.14 & \textbf{32.85} & 42.49 & OOM & OOM & \textbf{1143.98} & OOM & OOM  \\
     & 12288 & 190.40 & 238.77 & 278.84 & \textbf{65.10} & 99.12 & OOM & OOM & OOM & OOM & OOM  \\
     & 24576 & 381.68 & OOM & 588.15 & \textbf{130.76} & 186.94 & OOM & OOM & OOM & OOM & OOM  \\
     & 49152 & OOM & OOM & OOM & OOM & \textbf{403.38} & OOM & OOM & OOM & OOM & OOM  \\
     & 98304 & OOM & OOM & OOM & OOM & OOM & OOM & OOM & OOM & OOM & OOM  \\
    \hline
    \multirow{10}{*}{FP32}
     & 192 & 1.91 & 5.17 & 4.29 & \textbf{1.76} & 4.54 & \textbf{4.62} & 10.27 & 6.37 & 48.74 & \textbf{9.11}  \\
     & 384 & 2.19 & 5.37 & 5.57 & \textbf{1.91} & 4.78 & \textbf{8.85} & 12.58 & 10.43 & 51.07 & \textbf{15.63}  \\
     & 768 & 3.13 & 5.49 & 7.33 & \textbf{2.45} & 5.26 & 21.08 & 25.28 & \textbf{20.08} & 61.40 & \textbf{28.95}  \\
     & 1536 & 6.24 & 7.87 & 12.22 & \textbf{3.45} & 6.12 & 46.04 & 55.34 & \textbf{38.99} & 110.43 & \textbf{63.41}  \\
     & 3072 & 12.51 & 16.63 & 21.69 & \textbf{5.49} & 8.23 & 93.44 & 110.10 & \textbf{75.57} & 217.39 & \textbf{142.86}  \\
     & 6144 & 25.17 & 29.60 & 40.39 & \textbf{10.31} & 15.64 & 186.86 & 220.43 & \textbf{148.14} & OOM & \textbf{339.70}  \\
     & 12288 & 50.52 & 58.01 & 97.20 & \textbf{20.08} & 25.82 & OOM & OOM & \textbf{296.33} & OOM & OOM  \\
     & 24576 & 102.33 & 140.70 & 196.08 & \textbf{40.59} & 75.22 & OOM & OOM & OOM & OOM & OOM  \\
     & 49152 & OOM & OOM & 386.50 & \textbf{83.72} & 170.30 & OOM & OOM & OOM & OOM & OOM  \\
     & 98304 & OOM & OOM & OOM & OOM & \textbf{362.81} & OOM & OOM & OOM & OOM & OOM  \\
    \end{tabular}
    }
    \caption{The molecular dynamics performance (ms/step) of DPA-1 (L=0), DPA-1 (L=2), and DPA-2 models in double (FP64) and single (FP32) precisions in TensorFlow (TF), PyTorch (PT), and JAX backends on a single 24~GB NVIDIA 4090 GPU card.
    The water system with different numbers of atoms is used for simulations.
    ``TFc'' and ``PTc'' donate to compressed models in TensorFlow and PyTorch backends.
    ``OOM'' donates to out-of-memory (GPU memory exceeds 24~GB).
    The best performance for each model is highlighted in bold.
    }
    \label{tab:benchmark}
\end{table}

\begin{figure}[H]
    \includegraphics[width=\linewidth]{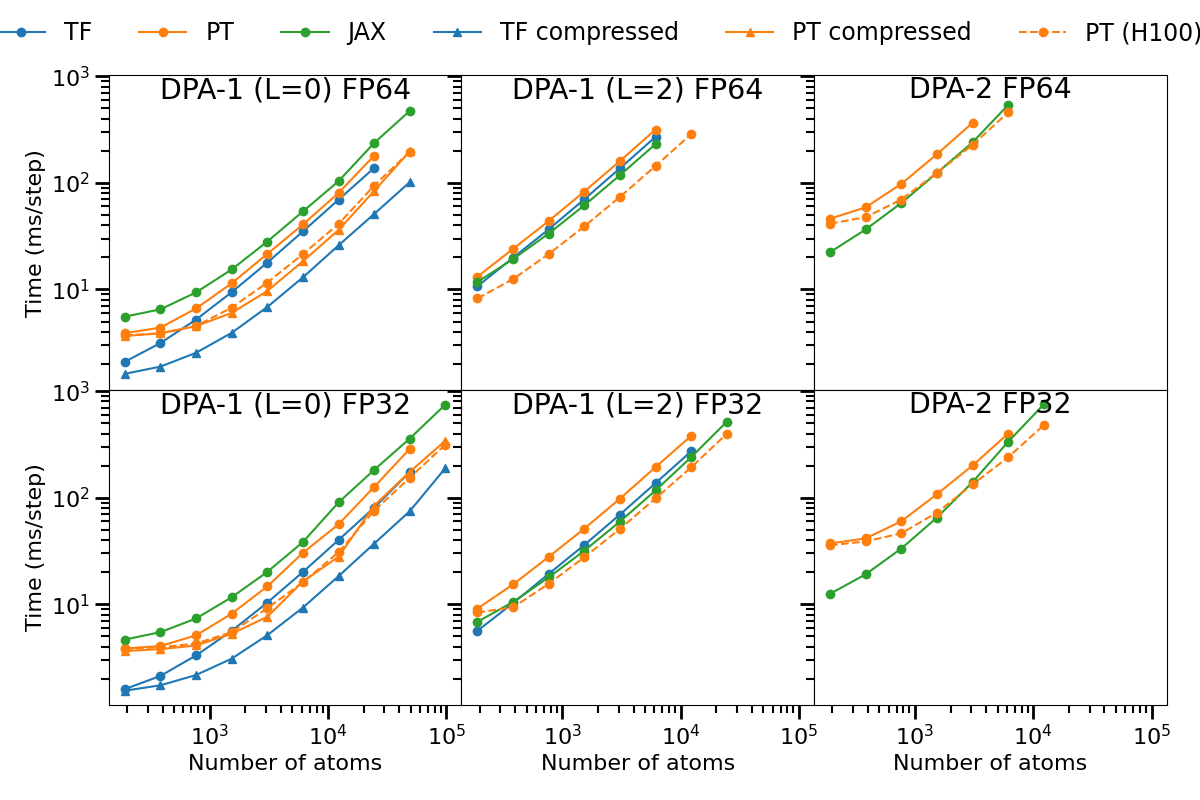}
    \centering
    \caption{The molecular dynamics performance (ms/step) of DPA-1 (L=0), DPA-1 (L=2), and DPA-2 models in double (FP64) and single (FP32) precisions in TensorFlow (TF), PyTorch (PT), and JAX backends on a single 40~GB NVIDIA A800 GPU card.
    The water system with different numbers of atoms is used for simulations.
    }
    \label{fig:a800}
\end{figure}

\begin{figure}[H]
    \includegraphics[width=\linewidth]{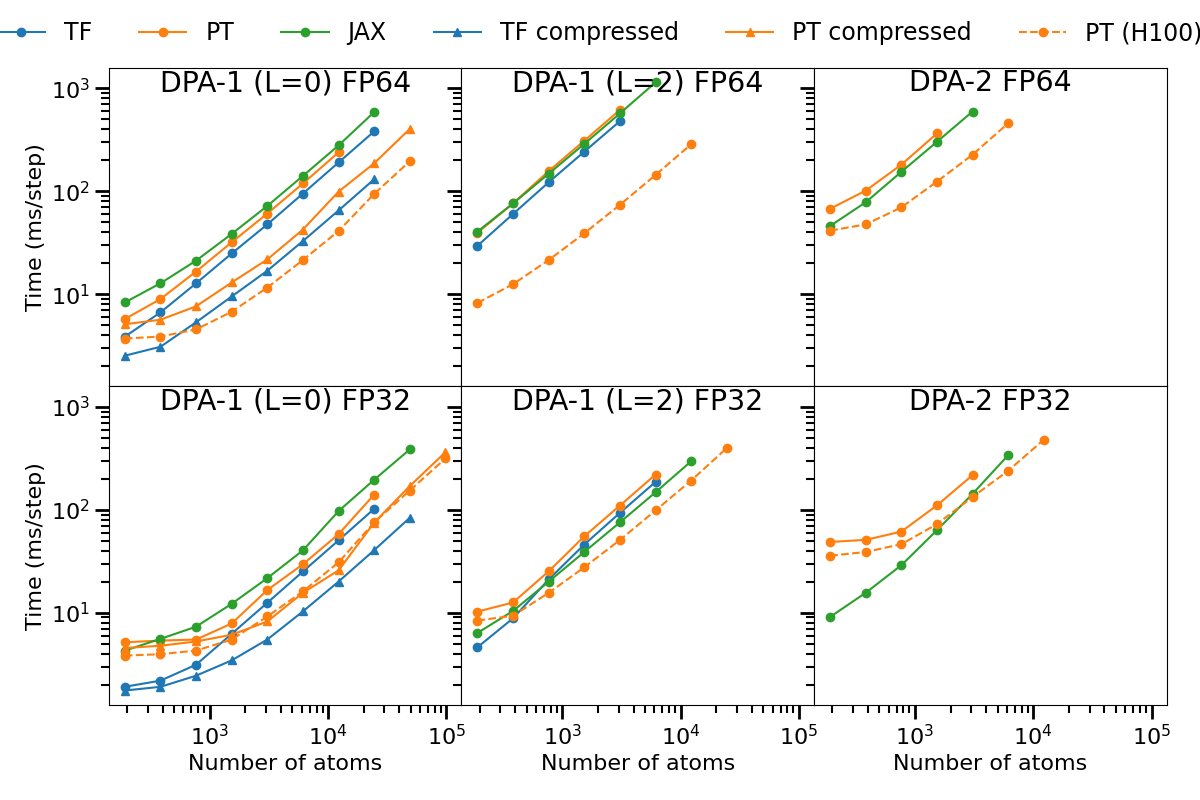}
    \centering
    \caption{The molecular dynamics performance (ms/step) of DPA-1 (L=0), DPA-1 (L=2), and DPA-2 models in double (FP64) and single (FP32) precisions in TensorFlow (TF), PyTorch (PT), and JAX backends on a single 24~GB NVIDIA 4090 GPU card.
    The water system with different numbers of atoms is used for simulations.
    }
    \label{fig:4090}
\end{figure}

\newpage

\section{Example of Using Large Language Models for Code Conversion between Backends}

To demonstrate how large language models can efficiently convert codes between backends, we provide an example where the DeepSeek R1 is used to convert code for the DeepPot-SE descriptor.

Prompt: Convert to PyTorch implementation.

Input code:

\begin{lstlisting}[language=Python]
def call(
    self,
    coord_ext,
    atype_ext,
    nlist,
    mapping: Optional[np.ndarray] = None,
):
    xp = array_api_compat.array_namespace(coord_ext, atype_ext, nlist)
    input_dtype = coord_ext.dtype
    # nf x nloc x nnei x 4
    rr, diff, ww = self.env_mat.call(
        coord_ext, atype_ext, nlist, self.davg, self.dstd
    )
    nf, nloc, nnei, _ = rr.shape
    sec = self.sel_cumsum

    ng = self.neuron[-1]
    gr = xp.zeros([nf * nloc, ng, 4], dtype=self.dstd.dtype)
    exclude_mask = self.emask.build_type_exclude_mask(nlist, atype_ext)
    # merge nf and nloc axis, so for type_one_side == False,
    # we don't require atype is the same in all frames
    exclude_mask = xp.reshape(exclude_mask, (nf * nloc, nnei))
    rr = xp.reshape(rr, (nf * nloc, nnei, 4))
    rr = xp.astype(rr, self.dstd.dtype)

    for embedding_idx in itertools.product(
        range(self.ntypes), repeat=self.embeddings.ndim
    ):
        (tt,) = embedding_idx
        mm = exclude_mask[:, sec[tt] : sec[tt + 1]]
        tr = rr[:, sec[tt] : sec[tt + 1], :]
        tr = tr * xp.astype(mm[:, :, None], tr.dtype)
        ss = tr[..., 0:1]
        gg = self.cal_g(ss, embedding_idx)
        # gr_tmp = xp.einsum("lni,lnj->lij", gg, tr)
        gr_tmp = xp.sum(gg[:, :, :, None] * tr[:, :, None, :], axis=1)
        gr += gr_tmp
    gr = xp.reshape(gr, (nf, nloc, ng, 4))
    # nf x nloc x ng x 4
    gr /= self.nnei
    gr1 = gr[:, :, : self.axis_neuron, :]
    # nf x nloc x ng x ng1
    # grrg = xp.einsum("flid,fljd->flij", gr, gr1)
    grrg = xp.sum(gr[:, :, :, None, :] * gr1[:, :, None, :, :], axis=4)
    # nf x nloc x (ng x ng1)
    grrg = xp.reshape(grrg, (nf, nloc, ng * self.axis_neuron))
    return grrg, gr[..., 1:], None, None, ww
\end{lstlisting}

Output code:

\begin{lstlisting}[language=Python]
import torch
import itertools

def call(
    self,
    coord_ext: torch.Tensor,
    atype_ext: torch.Tensor,
    nlist: torch.Tensor,
    mapping: Optional[torch.Tensor] = None,
):
    input_dtype = coord_ext.dtype  # Preserve input dtype for reference

    # Compute environment matrices using PyTorch operations
    rr, diff, ww = self.env_mat.call(
        coord_ext, atype_ext, nlist, self.davg, self.dstd
    )
    nf, nloc, nnei, _ = rr.shape
    sec = self.sel_cumsum  # Cumulative sum of selections

    ng = self.neuron[-1]
    # Initialize output tensor on the same device as input
    gr = torch.zeros(
        (nf * nloc, ng, 4),
        dtype=self.dstd.dtype,
        device=coord_ext.device
    )

    # Build and reshape exclusion mask
    exclude_mask = self.emask.build_type_exclude_mask(nlist, atype_ext)
    exclude_mask = exclude_mask.reshape(nf * nloc, nnei)

    # Reshape and cast radial components
    rr = rr.reshape(nf * nloc, nnei, 4).to(dtype=self.dstd.dtype)

    # Process each atom type embedding
    for embedding_idx in itertools.product(
        range(self.ntypes),
        repeat=self.embeddings.ndim
    ):
        (tt,) = embedding_idx  # Unpack single-element tuple
        start_idx = sec[tt]
        end_idx = sec[tt + 1]

        # Slice relevant components for current type
        type_mask = exclude_mask[:, start_idx:end_idx]
        type_rr = rr[:, start_idx:end_idx, :]

        # Apply mask and compute scalar features
        type_rr = type_rr * type_mask.unsqueeze(-1).to(dtype=type_rr.dtype)
        scalar_features = type_rr[..., 0:1]

        # Compute radial gate and aggregate contributions
        radial_gate = self.cal_g(scalar_features, embedding_idx)
        gr_tmp = (radial_gate.unsqueeze(-1) * type_rr.unsqueeze(2)).sum(dim=1)
        gr += gr_tmp

    # Reshape and normalize aggregated features
    gr = gr.reshape(nf, nloc, ng, 4) / self.nnei

    # Compute final feature products
    gr1 = gr[:, :, :self.axis_neuron, :]
    grrg = (gr.unsqueeze(3) * gr1.unsqueeze(2)).sum(dim=-1)
    grrg = grrg.reshape(nf, nloc, -1)

    return grrg, gr[..., 1:], None, None, ww
\end{lstlisting}